\documentclass[twocolumn]{article}
\usepackage[margin=20mm]{geometry}
\usepackage{authblk}
\usepackage{cite}
\usepackage{lineno}
\usepackage{siunitx}
\usepackage{amsmath}
\usepackage{amsfonts}
\usepackage{amssymb}
\usepackage{graphicx}
\usepackage{hyperref}
\usepackage{cancel}
\usepackage{multirow}
\usepackage{changepage}
\usepackage{float}
\usepackage[skip=10pt]{caption}
\hypersetup{
    colorlinks=true,
    linkcolor=blue,
    filecolor=magenta,      
    urlcolor=cyan,
    }

\providecommand{\keywords}[1]
{
  \small	
  \textbf{\textit{Keywords---}} #1
}

\title{
Reconstruction of three-dimensional fluid stress field \\
via photoelasticity using \\
physics-informed convolutional encoder-decoder}

\author[1]{Daichi Igarashi}
\author[1]{Shunsuke Kumagai}
\author[1,2]{Yuto Yokoyama}
\author[1]{Yee Jingzu}
\author[3]{Masanobu Horie}
\author[1]{Yoshiyuki Tagawa}

\affil[1]{Department of Mechanical Systems Engineering, Tokyo University of Agriculture and Technology, Koganei, Tokyo 1848588, Japan}
\affil[2]{Okinawa Institute of Science and Technology, Kunigami-gun, Okinawa 9040495, Japan}
\affil[3]{RICOS Co. Ltd., Chiyoda-ku, Tokyo 1000005, Japan}

\date{\today}

\begin{document}

\twocolumn[

\maketitle

\begin{abstract}
Measuring stress fields in fluids and soft materials is crucial in various fields such as mechanical engineering, medicine, and bioengineering.
However, conventional methods that calculate stress fields from velocity fields struggle to measure complex fluids where the stress constitutive equation is unknown.
To address this, we propose a novel approach that combines photoelastic measurements---which can non-invasively visualize internal stresses---with machine learning to measure stress fields.
The machine learning model, which we named physics-informed convolutional encoder-decoder (PICED), integrates a convolutional neural network (CNN)-based encoder-decoder model with a physics-informed neural network (PINN).
Using this approach, three-dimensional stress fields can be predicted with high accuracy for multiple interpolated data points in a rectangular channel flow.
\end{abstract}

\vspace{3mm}

\keywords{
machine learning,
data-driven modeling,
physics-informed neural networks,
photoelasticity,
}

\vspace{8mm}
]

\section{Introduction}
The measurement of the nonlinear dynamics of fluids and soft materials has wide-ranging applications in various fields, such as mechanical engineering, medicine, and bioengineering.
Nonlinear dynamics involve complex relationships between deformation and stress, and stress measurement provides insights into phenomena such as cerebral aneurysm formation \cite{Meng1254,van_ooij_characterization_2015}, cellular mechanical responses \cite{traub1998laminar,tzima2005mechanosensory}, and the mechanical basis of biochemical changes in cancer cells \cite{mitchell2013computational,lee2017fluid}.
In engineering, understanding stress distribution is critical for advancing needle-free injectors \cite{tagawa2012highly,igarashi2024effects} and improving the structural analysis of soft robots \cite{bartlett20153d,vikas2016design}.

In fluid dynamic interactions such as those described above, the stress field is typically derived from velocity measurements using a constitutive equation or pressure sensors.
However, these methods face limitations, particularly for complex fluids lacking known constitutive equations or in situations when it is crucial to avoid disturbances to the flow. 
To address such limitations, this study focuses on photoelastic measurement, a non-contact method for directly measuring stress fields within materials.
Photoelastic measurement is an optical technique that utilizes the property of transparent materials, where the refractive index changes when stress is applied \cite{drucker1940stress,aben2000integrated,ramesh2016digital}.
While this technique has been widely applied to solid materials, such as glass, recent studies have extended its use to fluids and soft materials \cite{lane2021optical,lane2022birefringent,calabrese2022alignment,full1995maximum,miyazaki2021dynamic,yokoyama2023integrated,nakamine_flow_2024,worby_examination_2024}.
When using a photoelastic fluid containing high-aspect-ratio particles, the particles align in the direction of stress and exhibit an optical phase difference $\Delta$, which corresponds to the magnitude of the stress.
This phase difference and particle orientation $\phi$ are related to the difference and direction of the principal stresses \cite{prabhakaran1975stress,janeschitz2012polymer}, enabling the direct, non-contact measurement of stress fields in fluids.

To determine the actual stress field in a fluid, it is necessary to reconstruct it from the photoelastic parameters (phase difference and orientation), which are obtained via a polarization camera.
A challenge arises because stress is a tensor with up to six independent components at each point, while the photoelastic parameters provide only two values, which are integrated along the optical axis.
This discrepancy makes reconstruction difficult.
Previous photoelastic measurements have assumed axial symmetry to simplify the process by reducing the number of stress components \cite{errapart2011,anton_discrete_2008}.
Although several studies have calculated phase difference and orientation from a 3D stress field \cite{aben1997photoelastic,ober2011spatially,kim2017monitoring}, to the authors' knowledge, no research has reconstructed a 3D fluid stress field from these parameters without assuming axial symmetry.

As mentioned above, reconstructing the 3D stress tensor field in a fluid from phase difference and orientation data requires complex matrix calculations; however, this involves a nonlinear problem with no theoretical solution.
Therefore, the present study focuses on machine learning, which can handle nonlinear problems.
In recent years, machine learning has been integrated with fluid mechanics, especially in physics-informed neural networks (PINNs) \cite{karniadakis2021physics,cai2021physics,jin2021nsfnets}, which satisfy the governing physical equations of fluid mechanics.
These approaches have been used to predict pressure distributions \cite{kissas2020machine} and fluid behavior in complex fluids \cite{mahmoudabadbozchelou2022nn}, but none have focused on predicting stress fields.
Some studies have also enhanced model generalization by incorporating numerical differentiation \cite{horie2021isometric,horie2022physics,horie2024graph}.

This study's goal is to reconstruct 3D stress fields from 2D phase difference and orientation images using a machine learning model called physics-informed convolutional encoder-decoder (PICED), which combines a CNN-based encoder-decoder and a PINN.
To verify the model before applying it to more complex flow paths, this study focuses on a simpler system---a rectangular channel.

Section 2 explains the machine learning architecture and the method for obtaining the data used for the input and output images.
Section 3 describes the prediction results of the three-dimensional stress field, the principal stress distribution, and the element analysis. 
Finally, in Section 4, the conclusions of this study are presented, and future prospects are discussed.
\section{Methodology}
This section describes the method for obtaining 3D stress distributions from integrated images of phase difference and orientation, which are acquired through polarization measurements.
In this study, we reconstructed the stress field for a Newtonian fluid in laminar flow within a rectangular channel.
The input images were obtained using the experimental methods described in Sections 2.1.1 and 2.1.2, while the output images were derived from the theoretical solutions detailed in Section 2.2. Finally, Section 2.3 illustrates the structure of the machine learning model---specifically, a convolutional neural network (CNN) which is generally used and the PICED method developed in this study---and provides a comprehensive explanation of how the three-dimensional stress field is obtained.

\subsection{Input: Two-dimensional integrated
images of phase difference and orientation}
This section explains the methodology for acquiring input images from experiments.

\subsubsection{Measurement principle}
\begin{figure}[!b]
\centering
\includegraphics[width=1.0\columnwidth]{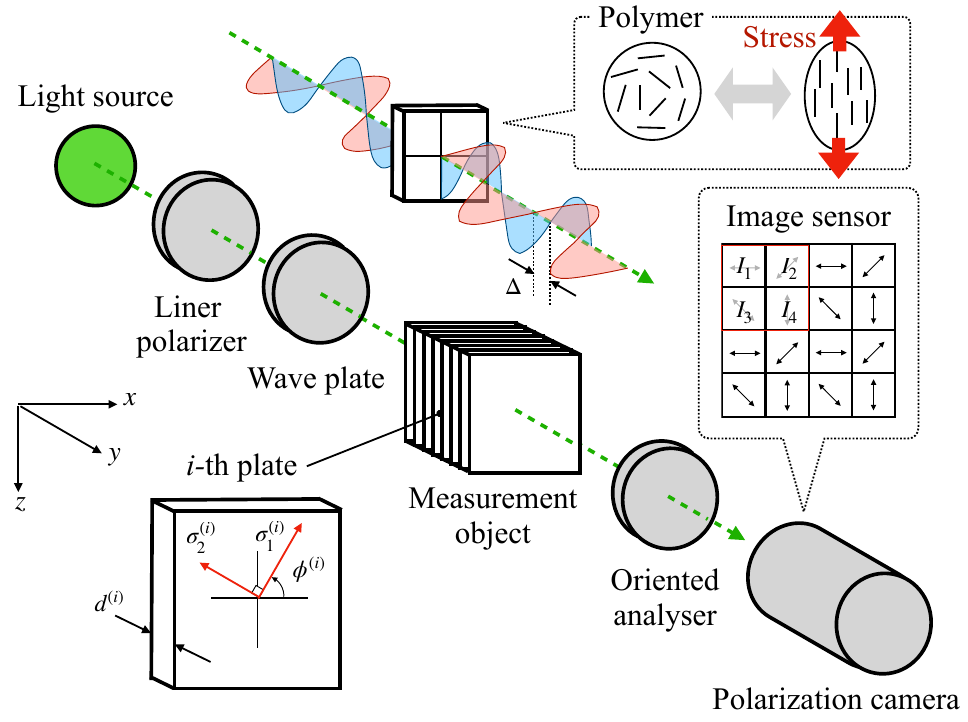}
\caption{Schematic drawing of the theory of the photoelastic method.}
\label{fig:photo}
\end{figure}

The optical system is illustrated in Figure \ref{fig:photo}.
A photoelastic material is positioned along the optical axis between a polarization camera and a circularly polarized light source.
The refractive index varies according to the direction of polarization when the photoelastic material is subjected to stress.
Since the speed of light is inversely proportional to the refractive index, an optical phase difference occurs when light passes through two orthogonal polarizers \cite{bass1995handbook}.
This phenomenon, whereby two refracted rays appear when light passes through a material, is known as birefringence.
Optically anisotropic particles, such as cellulose nanocrystal (CNC), exhibit birefringent properties \cite{nakamine_flow_2024,worby_examination_2024,lane2022birefringent}.
These particles are randomly oriented when no stress is applied but are aligned in a uniform direction under stress.
However, since the distortion caused by stress is an elastic deformation, birefringence disappears once the external force is removed.

The monochromatic light generated by the light source is converted to circularly polarized light by passing it through a linear polarizer and a quarter-wave plate.
Through circular polarization, the light exhibits uniform vibrations in all directions perpendicular to its propagation.
The transmitted light---oriented because of the birefringence of the photoelastic object---passes through a linear polarizer positioned between the camera and the photoelastic object.
This linear polarizer is rotated to four angles---$0^\circ$, $45^\circ$, $90^\circ$, and $135^\circ$---allowing the four polarizers of the polarizing camera to capture the brightness values of the transmitted light.
Let these brightness values be denoted as $I_1, I_2, I_3, I_4$; the phase difference $\Delta$ and the principal stress direction (orientation) $\phi$ can then be expressed as follows \cite{onuma2014development}:

\begin{equation} 
\label{eq:Delta_I}
\Delta = (\frac{\Lambda}{2\pi})\sin^{-1} \frac{\sqrt{(I_3-I_1)^2+(I_2-I_4)^2}}{I_1+I_2+I_3+I_4},
\end{equation}
\begin{equation} 
\label{eq:phi_I}
\phi = \frac{1}{2}\tan^{-1}\frac{I_3-I_1}{I_2-I_4}.
\end{equation}

\noindent Here, $\Lambda$ denotes the wavelength of the light source.
Furthermore, since an optoelastic material can be regarded as comprising multiple sliced plates, the values of phase difference and orientation represent the cumulative sums of the stress values applied to each individual plate.
These cumulative values are expressed by the following equation \cite{nakamine_flow_2024}:

\begin{align}
\label{eq:delta_3D}
V_1	\equiv \Delta \cos2\phi &= \int \left\{ C_1 (\sigma_{yy} - \sigma_{xx}) \right. \nonumber \\ 
&\quad + C_2 \left[ (\sigma_{yy} + \sigma_{xx})(\sigma_{yy} - \sigma_{xx}) \right. \nonumber \\ 
&\quad \left. \left. + \sigma_{yz}^2 - \sigma_{xz}^2 \right] \right\} \, dz,
\end{align}

\begin{align}
\label{eq:phi_3D}
V_2	\equiv \Delta \sin2\phi &= \int \left\{ 2C_1 \sigma_{xy} + C_2 \left[ 2(\sigma_{yy} + \sigma_{xx})\sigma_{xy} \right. \right. \nonumber \\ 
&\quad \left. \left. + 2\sigma_{yz}\sigma_{xz} \right] \right\} \, dz.
\end{align}

\noindent Here, $C_1$ and $C_2$ represent the stress-optical coefficients.
These equations lead to

\begin{equation} 
\label{eq:Delta_phi}
\Delta = \sqrt{V_1^2 + V_2^2}, \quad \phi = \frac{1}{2} \tan^{-1} \frac{V_2}{V_1}.
\end{equation}

It is evident from Equations (\ref{eq:delta_3D})--(\ref{eq:Delta_phi}) that the values of phase difference and orientation represent the integrated values of stress applied to multiple plates.
Consequently, the reconstruction of the 3D stress distribution from these integrated values poses a nonlinear problem.
In our study, this nonlinear problem is addressed using our proposed machine learning model, which is described in Section 2.3.

\subsubsection{Experimental setup and conditions}
This section describes the experimental methods for measuring phase difference and orientation in flow within a rectangular channel.
Nakamine et al. \cite{nakamine_flow_2024} conducted polarization measurements in a rectangular channel and developed a theory to obtain the distribution of integrated phase difference and orientation from the theoretical stress field in the fluid.
However, deriving the 3D stress distribution from the integrated values of phase difference and orientation leads to an indeterminate equation with non-unique solutions.
Thus, we attempt a 3D reconstruction of this rectangular channel using machine learning.

\begin{figure}[htbp]
\centering
\includegraphics[width=0.8\columnwidth]{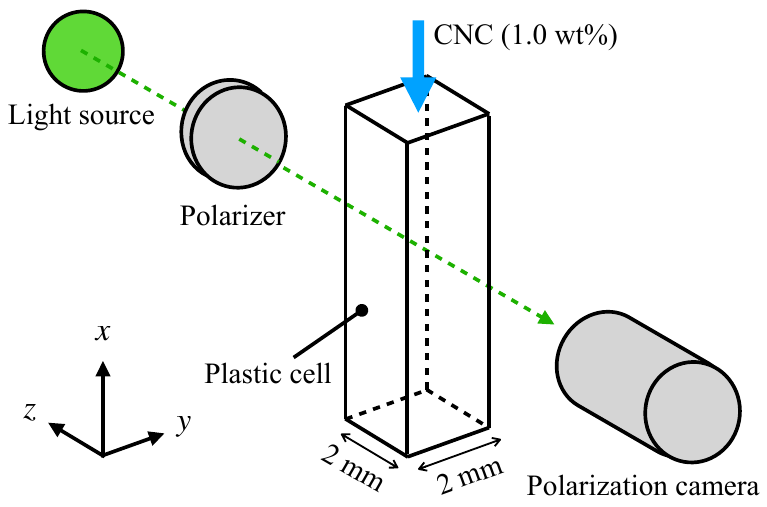}
\caption{Experimental setup for capturing the flow-induced birefringence in a rectangular channel.}
\label{fig:expsetup}
\end{figure}

Figure \ref{fig:expsetup} illustrates the setup for conducting experiments similar to those in the aforementioned research.
The light emitted from the light source (520 nm, SLG-55-G, REVOX Inc.) passes through a circular polarizer and the measurement target before entering a high-speed polarizing camera (CRYSTA PI-1P, Photron Ltd., frame rate: 250 f.p.s.).
The measurement target comprises a rectangular plastic channel with a cross-section of 2×2 mm.
We used a photoelastic fluid, CNC suspension (1.0 wt\%), as the working fluid.
This configuration enables polarization measurements based on the theory explained in the previous section.
In this experiment, the flow rate $Q$ was varied, and data were collected under nine conditions: $Q=5,10,20,30,40,50,60,70,80$ mL/min.
For each flow rate, 100 images of phase difference and orientation were obtained, and time averaging was performed to derive five datasets from every 20 images.

\subsubsection{Experimental data}

\begin{figure}[!t]
\centering
\includegraphics[width=1.0\columnwidth]{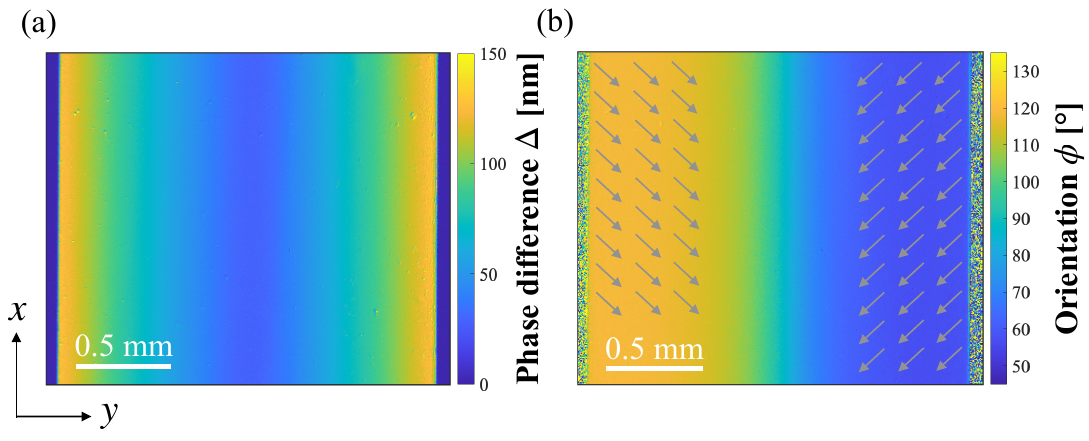}
\caption{Example images of (a) phase difference and (b) orientation captured by the polarization camera for the flow in the rectangular channel ($Q=40$ mL/min).}
\label{fig:expresults}
\end{figure}

\begin{figure}[!t]
\centering
\includegraphics[width=1.0\columnwidth]{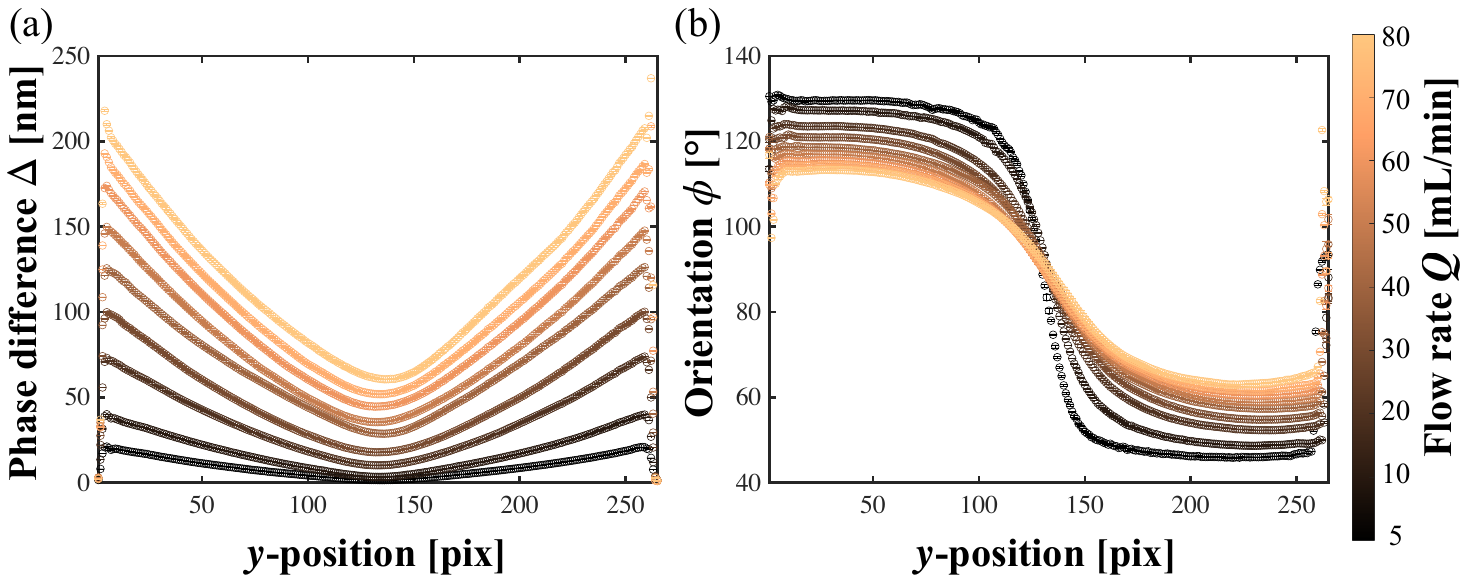}
\caption{Plots of (a) phase difference and (b) orientation with respect to the $y$-position for each flow rate.}
\label{fig:expresultsgraph}
\end{figure}

This section describes the input images obtained from the experiments and the image processing performed to prepare them as input for machine learning.
Figure \ref{fig:expresults} a and b illustrates the distributions of phase difference and orientation acquired from the experiments for the case in which  $Q=40$ mL/min.
Additionally, Figure \ref{fig:expresultsgraph}a and b plots the averaged values in the flow direction against the $y$-position for all flow rates.
The phase difference exhibits a symmetric distribution with respect to the center of the flow channel, reaching its minimum at the center and its maximum near the channel's walls.
As the flow rate increases, so does the absolute value of the phase difference.
The magnitude of the phase difference corresponds to the absolute value of the stress.
The orientation angle reaches its maximum near the left wall, its minimum near the right wall, and is 90° at the channel center, forming a point-symmetric distribution with respect to the horizontal axis.
This distribution shows that shear stresses dominate near the wall.
While the data at locations in contact with the walls deviate significantly from the trends near the walls, these values were used without modification in this analysis because of the difficulty in recognizing the wall surface.
The implications of using these values without modification are discussed in Section 3.1.

All images were resized from 340×424 pixels to 64×64 pixels using bicubic interpolation to reduce the machine learning computation time and facilitate shape restoration in the decoder.
In bicubic interpolation, the position on the original image corresponding to each pixel in the resized image is calculated.
Then, the values of the 16 neighboring pixels (4×4) around that position are weighted based on their distance, and the new pixel value is computed as a weighted average.
The test data comprised five datasets each for $Q=20, 40,$ and $60$ mL/min, while the remainder were utilized as training data.
To increase the amount of data, we utilized the symmetry of the phenomenon in the vertical direction and included vertically flipped versions of each dataset as training data.
As a result, 60 training datasets were prepared, comprising 10 datasets for each of the six flow rates.

\subsection{Output: Three-dimensional stress and velocity fields}

In this section, we explain the method for obtaining the 3D stress and velocity fields in flow within a rectangular channel.
The stress distribution, which serves as the ground truth data for machine learning, is derived from the theoretical solution for laminar flow of a Newtonian fluid in a rectangular channel, expressed by the following equations \cite{delplace2018laminar}, \cite{nagamoto2006laminar}:

\footnotesize
\begin{equation}
\label{eq:rectheory}
u_x(y,z)=\frac{64Q}{wh\pi^3}\sum^{\infty}_{n=0}\frac{(-1)^n}{(2n+1)^3}\Biggl[1-\frac{\cosh{\frac{(2n+1)\pi y}{2h}}}{\cosh{\frac{(2n+1)\pi w}{2h}}}\Biggl]\cos{\frac{(2n+1)\pi z}{2h}},
\end{equation}
\normalsize

\noindent where $w$ and $h$ are the channel width in the $y$-direction and channel thickness in the $z$-direction, respectively.

\begin{figure}[t!]
\centering
\includegraphics[width=1\columnwidth]{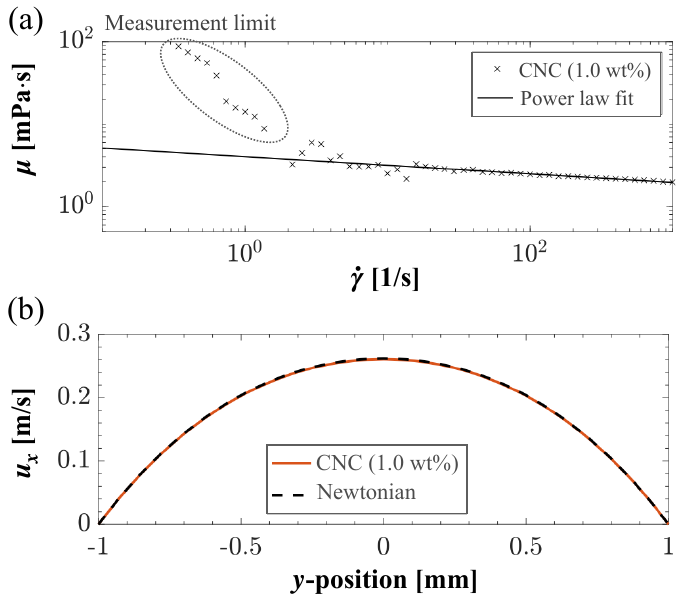}
\caption{(a) The relationship between shear rate and viscosity measured by a rheometer and (b) the velocity distribution of CNC and Newtonian fluids obtained by numerical simulations.}
\label{fig:shearcurve}
\end{figure}

Figure \ref{fig:shearcurve}a presents the relationship between the viscosity $\mu_{\rm{pl}}$ of the CNC (1.0 wt\%) used in this study and the shear rate, as measured by a rheometer (MCR302, Anton Paar Co. Ltd.).
From the figure, it can be observed that the viscosity exhibits a slight slope with respect to the shear rate, indicating shear-thinning behavior.
However, the region enclosed by the dotted line represents the rheometer's measurement limits, and we disregarded this area in our analysis.
Therefore, the same figure also presents the results of fitting the obtained data using the power law model, as described by the following equation:

\begin{equation}
\label{eq:shearcurve}
\mu_{\rm{pl}} = K \dot \gamma^{m-1}.
\end{equation}

\noindent The parameters $K$ and $m$ represent the fitting coefficients, with values obtained as $K \simeq 4.0$ and $m$ $\simeq 0.9$.
Using these values, we conducted numerical simulations for flow in a rectangular channel similar to the system used in this study.
The resulting distribution of the $x$-direction velocity $u_x$ against the $y$-position is shown in Figure \ref{fig:shearcurve}b.
The theoretical velocity distribution for the Newtonian fluid in this experimental setup, calculated using Equation (\ref{eq:rectheory}), is also presented in the same figure.
It is evident from the figure that both distributions exhibit approximately the same values.
On this basis, we can assume that the liquid used in this study behaves like a Newtonian fluid, and in this paper, we will utilize the theoretical solution described by Equation (\ref{eq:rectheory}) for Newtonian fluids.

Here, the components of the stress tensor in the rectangular channel flow satisfy the following equation:

\begin{equation}
\label{eq:sigma_3D}
\boldsymbol{\sigma}
=
\begin{bmatrix}
\sigma_{xx}&\sigma_{xy}&\sigma_{xz}\\
\sigma_{yx}&\sigma_{yy}&\sigma_{yz}\\
\sigma_{zx}&\sigma_{zy}&\sigma_{zz}
\end{bmatrix}
=
\begin{bmatrix}
-p&\sigma_{xy}&\sigma_{xz}\\
\sigma_{xy}&-p&0\\
\sigma_{xz}&0&-p
\end{bmatrix},
\end{equation}

\noindent where $p$ represents atmospheric pressure.
Therefore, the values that vary with each dataset are $\sigma_{xy}$ and $\sigma_{xz}$; thus, we only used these values as the output stresses.
The infinite shear rate viscosity of CNC was calculated from the rheometer measurements to be $\mu_{\rm{inf}}$ = 1.58.
Consequently, this viscosity value was utilized for the computation of the stress distribution.
Based on this, we conducted machine learning with the 2D distributions of phase difference and orientation as input and the 3D distribution of the stress tensor as output.

\subsection{Machine learning}
This section describes the machine learning methods for obtaining the stress field from phase difference and orientation in a 3D flow within a rectangular channel.
We compare the prediction results obtained using two models: a CNN and the PICED model developed in this study.

\subsubsection{Architecture of the CNN}

\begin{figure*}[htbp]
\centering
\includegraphics[width=2.0\columnwidth]{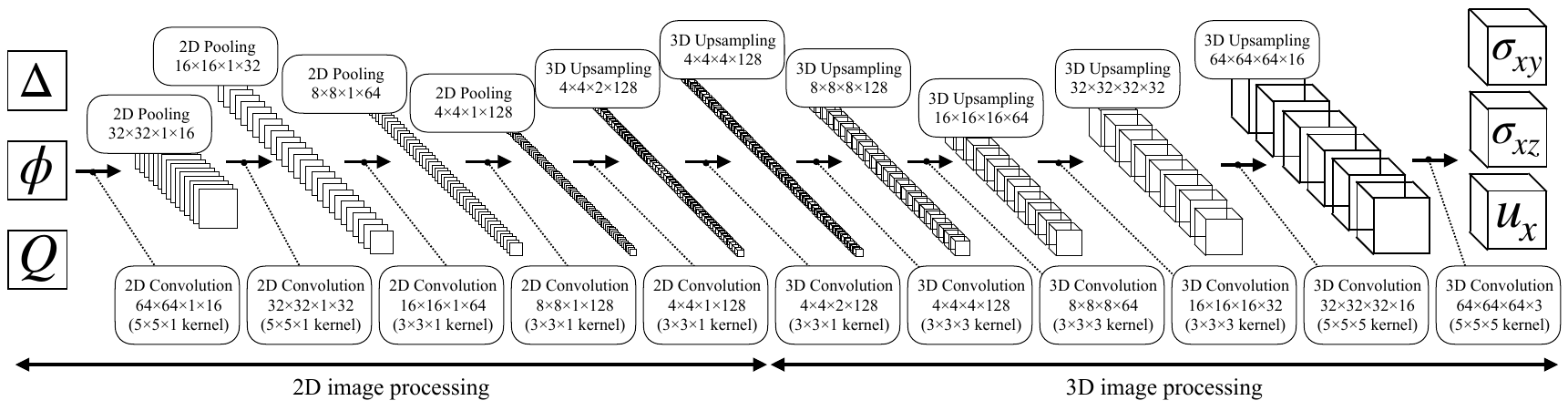}
\caption{Architecture of the convolutional encoder-decoder for the reconstruction of stress and velocity.}
\label{fig:MLmethod}
\end{figure*}

This section describes the architecture of the CNN used in this study for comparison purposes.
The CNN model is illustrated in Figure \ref{fig:MLmethod}.
It consists of an encoder for feature extraction and a decoder for returning to the original size.
During the encoding process, pooling operations are employed to aggregate multiple values in a specified region of the image into a single value, thereby reducing the overall size of the image.
In the decoder, upsampling is used to fill surrounding pixels with the same value as a single pixel.
Additionally, the convolutional encoder-decoder method utilizes convolutional operations alongside these processes, allowing for learning while maintaining the image shape.

In this paper, we employed a four-dimensional (4D) CNN that incorporates the three spatial components of velocity and stress components as channel dimensions.
From the 64×64 pixel images of phase difference and orientation, we performed max pooling and convolution four times, followed by seven convolutions and six upsampling operations to generate stress with the same size as the input image.
Note that because tuning results indicated the necessity for finer convolutions, especially when expanding in the depth direction (along the $z$-axis), we utilized a 21-layer deep neural network.
The activation function used in the intermediate layers was the ReLU (Rectified Linear Units) function \cite{agarap2018deep}, while the activation function in the output layer was the Sigmoid function \cite{cybenko1989approximation}.
During convolution, padding was applied to ensure that the size remains unchanged before and after the operation.
Furthermore, incorporating flow rate information into the input images in addition to phase difference and orientation resulted in improved accuracy.
Therefore, we included the flow rate as part of the input images.
The flow rate is represented in a 64×64 pixel image, where the value of the flow rate is filled in all pixels.

Since we used the mean squared error as the loss function $L_{\rm{data}}$, it can be expressed as follows:

\begin{equation}
\label{eq:loss}
L_{\rm{data}} = L_{\sigma xy} + L_{\sigma xz} + L_{ux},
\end{equation}

\noindent where

\begin{equation}
\label{eq:Ldata1}
L_{\sigma xy} = \frac{1}{N}\sum^N (\sigma_{xy}^{\rm{a}} - \sigma_{xy}^{\rm{p}})^2,
\end{equation}
\begin{equation}
\label{eq:Ldata2}
L_{\sigma xz} = \frac{1}{N}\sum^N (\sigma_{xz}^{\rm{a}} - \sigma_{xz}^{\rm{p}})^2,
\end{equation}
\begin{equation}
\label{eq:Ldata3}
L_{ux} = \frac{1}{N}\sum^N (u_x^{\rm{a}} - u_x^{\rm{p}})^2.
\end{equation}

\noindent The superscripts a and p represent the true and predicted values of the components, respectively, while $N$ denotes the number of data points. Furthermore, in the PICED model described later, the velocity $u_x$ in the $x$-direction is required by the physical equations, so velocity was also included as an output variable.
The true velocity distribution was obtained from the theoretical solution derived from Equation (\ref{eq:rectheory}).
For simplification in training, phase difference, orientation, flow rate, stress, and velocity were normalized by dividing each by the maximum value across all data.
However, the predicted values after training were transformed back to the original scale.

We used the Adam optimizer \cite{kingma2014adam} and performed parallel computation utilizing four GPUs (HPCT SR07V, NVIDIA RTX A6000) with data parallelism \cite{isard2007dryad}.
The computation time was approximately 170 ms per epoch.
Mini-batch learning \cite{li2014efficient} was conducted with a batch size of 16, and the validation set comprised 20\% of the total data.
To prevent overfitting, we employed early stopping \cite{prechelt2002early}, terminating the training if the minimum error was not updated within 500 epochs after reaching the minimum value.
The learning rate was set to 0.0001.
For reproducibility evaluation, training and testing were conducted five times using the same model.
The TensorFlow library was utilized for implementation.

\subsubsection{Architecture of the PICED}
\begin{figure}[t!]
\centering
\includegraphics[width=1.0\columnwidth]{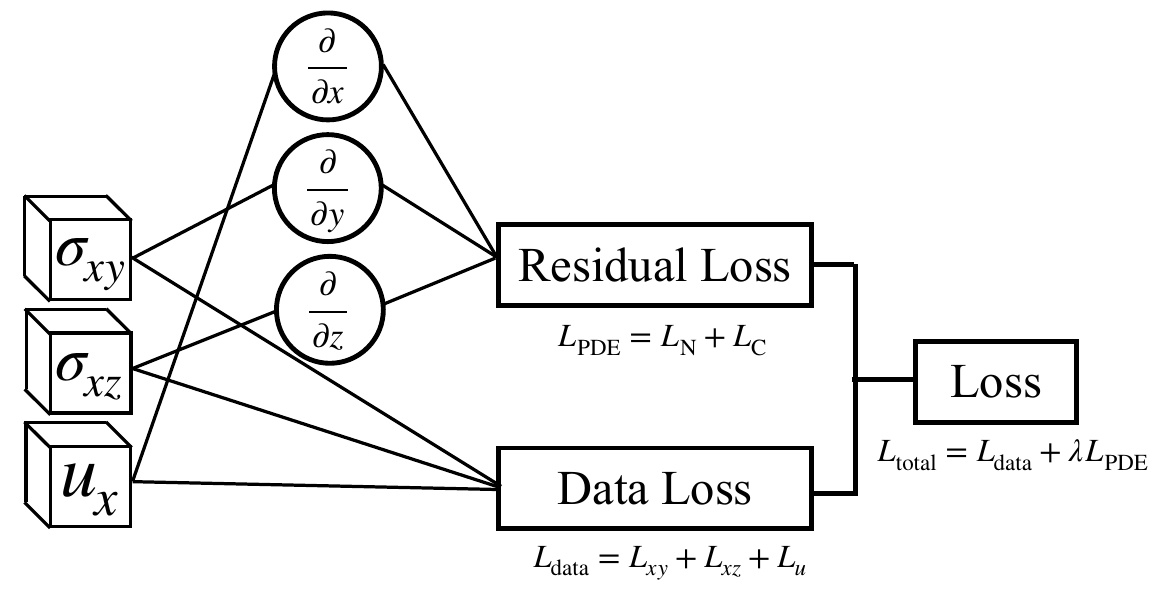}
\caption{Loss function of the physics-informed convolutional encoder-decoder (PICED).}
\label{fig:MLmethod_2}
\end{figure}

This section describes the structure of the PICED model developed in this study, using the architecture of the CNN from the previous section.
The PINN enables predictions that take physical phenomena into account by incorporating physical equations into the loss function.
Therefore, the loss function for the PINN can be expressed as follows:

\begin{equation}
\label{eq:loss_total}
L_{\rm{total}} = L_{\rm{data}} + \lambda L_{{\rm{PDE}}},
\end{equation}

\noindent where $L_{{\rm{PDE}}}$ represents the loss due to the physical equations, and $L_{\rm{total}}$ denotes the overall loss.
The PICED model incorporates physical equations into the loss function using the CNN from the previous section, as shown in Figure \ref{fig:MLmethod_2}.

For the physical equations, we utilized Cauchy's equation of motion and the continuity equation.
Cauchy's equation of motion was incorporated because it allows for the prediction of stress fields where the viscosity distribution is unknown, making it more versatile than deriving flow velocity and stress using viscous laws.
Consequently, $L_{{\rm{PDE}}}$ is expressed as follows:

\begin{equation}
\label{eq:loss_PDE}
L_{\rm{PDE}} = L_{\rm{N}} + \lambda' L_{\rm{C}},
\end{equation}

\noindent where

\begin{equation}
\label{eq:LN}
L_{\rm{N}} = \frac{1}{N}\sum^N \Bigr\{\frac{1}{\rho}\Bigl(\frac{\partial \sigma_{xy}^{\rm{p}}}{\partial y} + \frac{\partial \sigma_{xz}^{\rm{p}}}{\partial z} \Bigl) - u_x^{\rm{p}}\frac{\partial u_x^{\rm{p}}}{\partial x}\Bigr\}^2,
\end{equation}
\begin{equation}
\label{eq:LC}
L_{\rm{C}} = \frac{1}{N}\sum^N \Bigr(\frac{\partial u_x^{\rm{p}}}{\partial x}\Bigr)^2.
\end{equation}

\noindent For the data error, we used the mean squared error (MSE), similar to the process described in the previous section.

Additionally, as a result of parameter tuning, we set
$\lambda = 1.0$ in Equation (\ref{eq:loss_total}) and $\lambda' = 1.0 \times 10^{-2}$ in Equation (\ref{eq:loss_PDE}).
The three-point central difference method was employed for the differentiation operations in $L_{\rm{N}}$ and $L_{\rm{C}}$.
Furthermore, because of significant errors from differentiation operations near the walls of the computational domain, a region of one pixel from the wall was excluded from the calculation of $L_{\rm{PDE}}$.

\section{Results and Discussion}
In this section, we discuss the reconstruction of the stress field in the context of 3D flow, focusing on the flow within a rectangular channel.

\subsection{Prediction results of 3D stress fields}
\begin{figure}[!t]
\centering
\includegraphics[width=0.9\columnwidth]{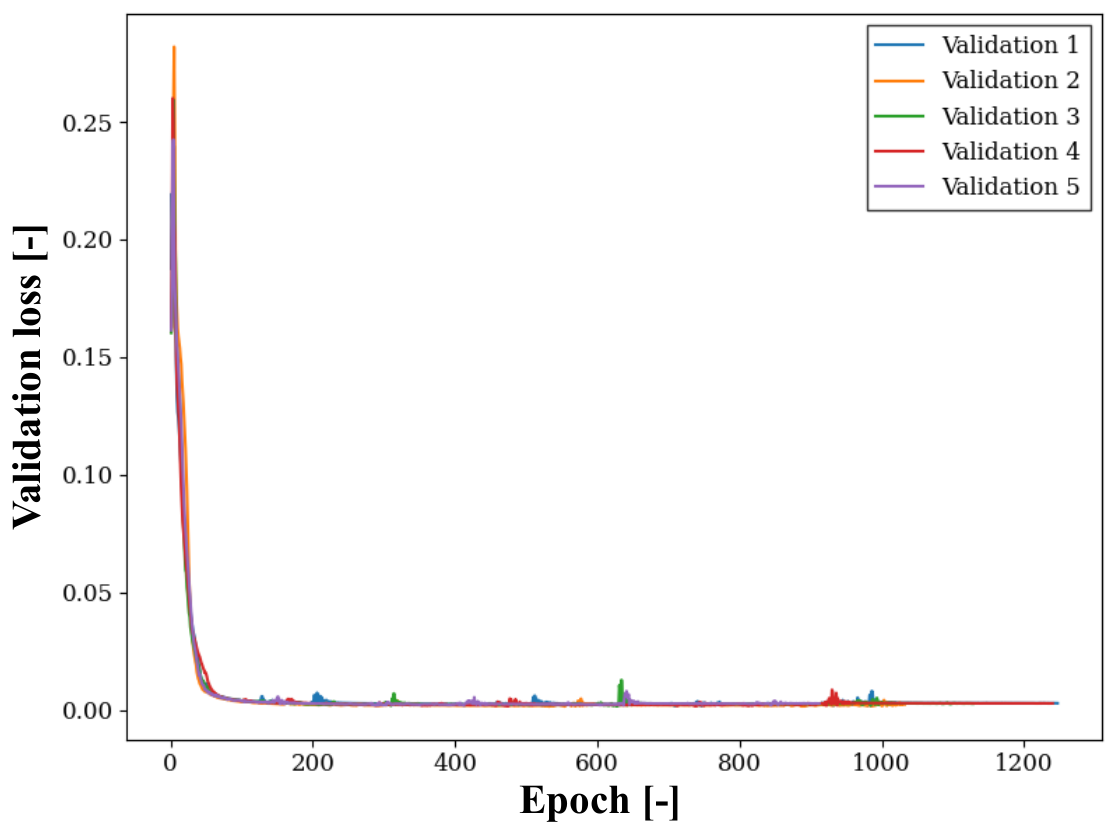}
\caption{Validation losses with respect to epochs for five validation tests.}
\label{fig:lossepoch}
\end{figure}

Here, we present and discuss the results of the reconstruction of the 3D stress field using the methods described in the previous sections.
First, Figure \ref{fig:lossepoch} demonstrates the progression of the error over epochs for the validation data.
Each line in the figure corresponds to the results from the five separate training sessions.
From the figure, it is evident that the loss decreases in all training sessions, indicating successful learning.
Therefore, we proceeded to test the model using the current learning framework.

\begin{figure*}[t!]
\centering
\includegraphics[width=1.75\columnwidth]{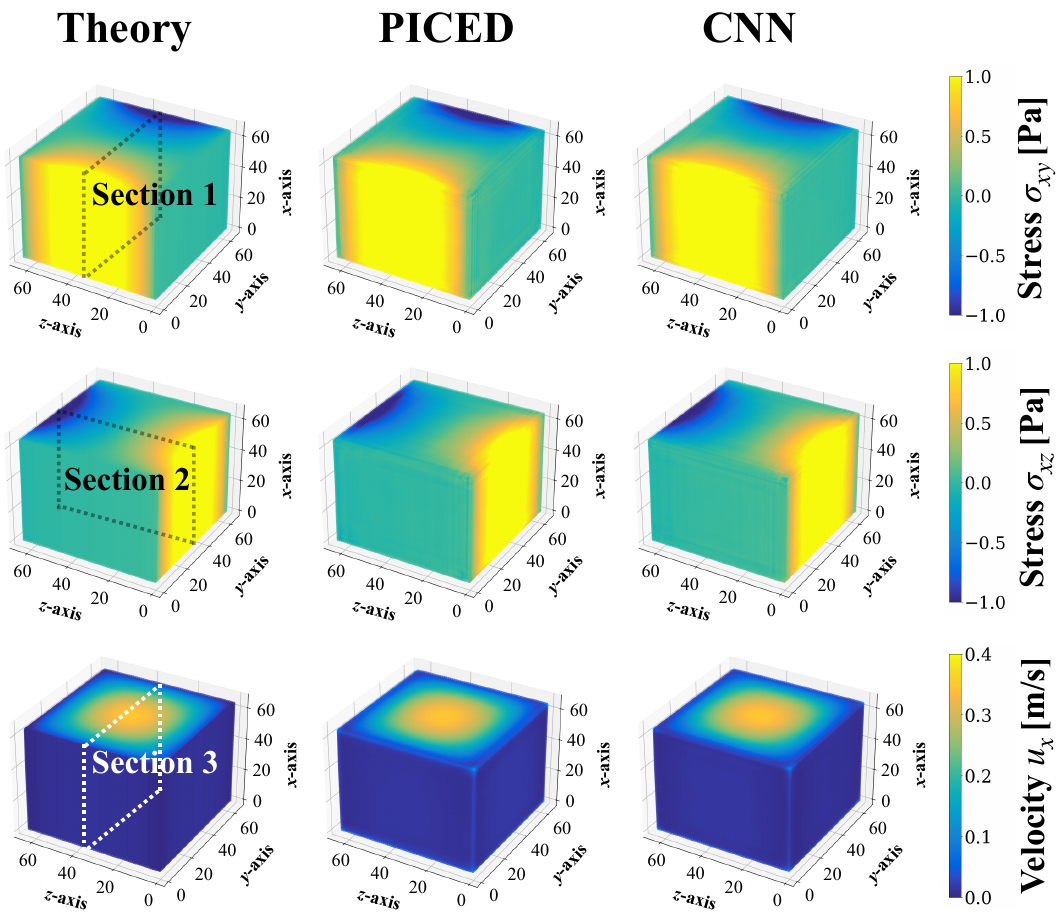}
\caption{3D distributions of stress and velocity: theoretical values, prediction values of PICED and CNN for $Q = 40$ mL/min.}
\label{fig:MLresults40}
\end{figure*}

Next, Figure \ref{fig:MLresults40} presents the results of the reconstruction of the stress and velocity fields for the test data at $Q=40$ mL/min.
In addition to the theoretical solution (Theory), which serves as the ground truth, we also include the reconstruction results using PICED, as well as the results obtained using only the MSE for the loss function (CNN) under the same conditions.
From the figure, it is evident that both PICED and CNN exhibit trends similar to those of the theoretical solution at the surfaces of the channel, and they also show comparable trends in the depth direction (the $z$-axis), which is not visible from the integrated images.

\begin{figure}[t!]
\centering
\includegraphics[width=1.0\columnwidth]{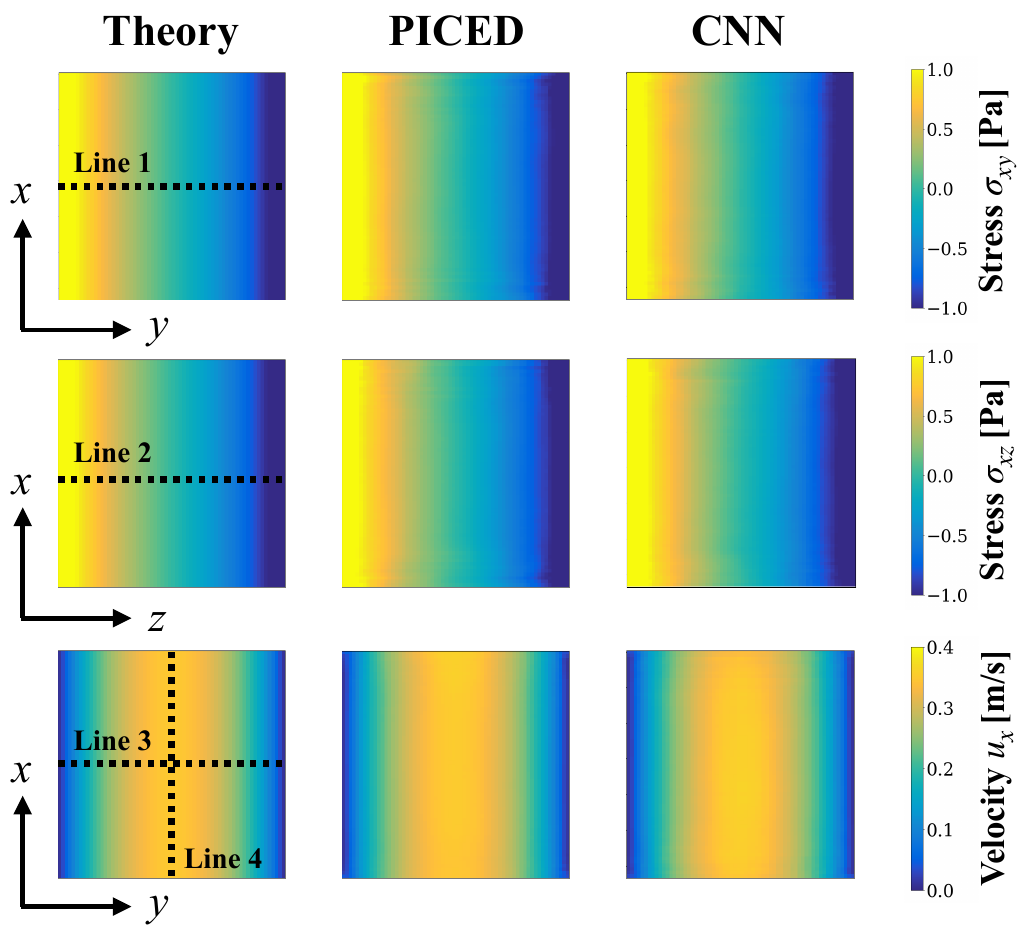}
\caption{2D distributions of stress and velocity: theoretical values, prediction values of PICED and CNN for sections 1--3 in Fig. 9 ($Q = 40$ mL/min).}
\label{fig:MLresults40danmen}
\end{figure}

\begin{figure}[htbp]
\centering
\includegraphics[width=0.74\columnwidth]{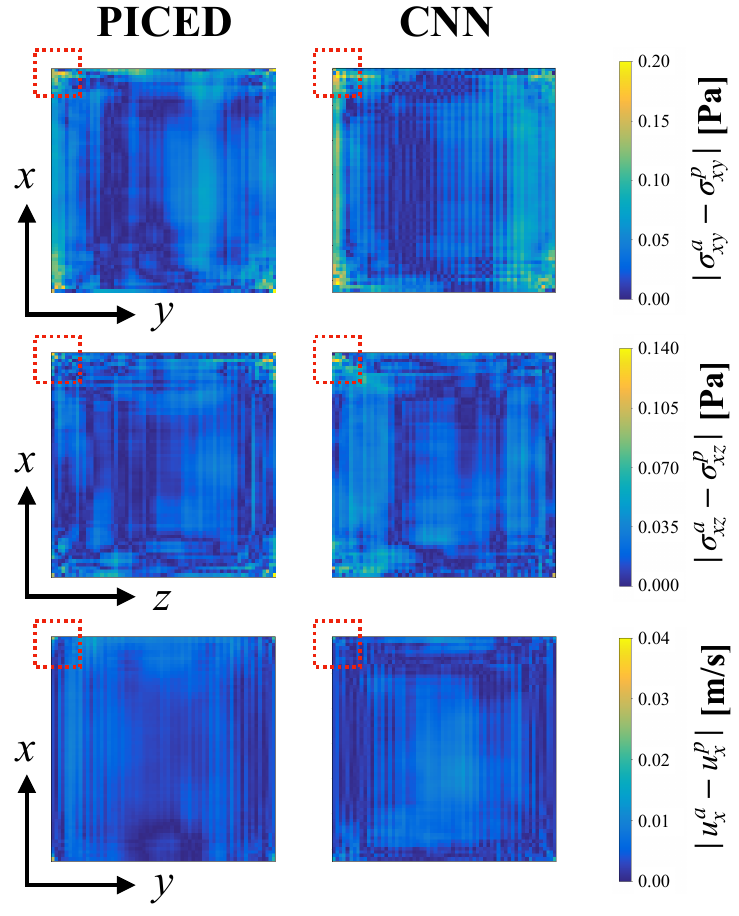}
\caption{2D absolute error distributions of stress and velocity for sections 1--3 in Fig. 9 ($Q = 40$ mL/min). The red dotted line indicates that the error is large near the boundary.}
\label{fig:MLresults40_absolute_error}
\end{figure}

To evaluate the predicted values inside the channel as well as at its surface, Figure \ref{fig:MLresults40danmen} presents the profiles of the three cross-sections depicted in Figure \ref{fig:MLresults40}.
The figure demonstrates that both PICED and CNN exhibit trends similar to those of the theoretical solution, not only at the surface but also inside the channel.
To accurately understand the errors relative to the theoretical solution, we calculated the absolute errors between the predicted values and the theoretical values, as shown in Figure \ref{fig:MLresults40_absolute_error}.
From this figure, it is evident that both PICED and CNN have larger errors near the boundaries.
This is likely due to the nature of convolutional processing, which makes predictions near the boundaries challenging, as well as the use of values outside the flow channel in the input images.
Therefore, to improve the prediction accuracy near the boundaries, it is necessary to apply boundary conditions to the physical equations and utilize segmentation techniques for more accurate recognition of the flow channel boundaries.

\begin{figure*}[htbp]
\centering
\includegraphics[width=2.0\columnwidth]{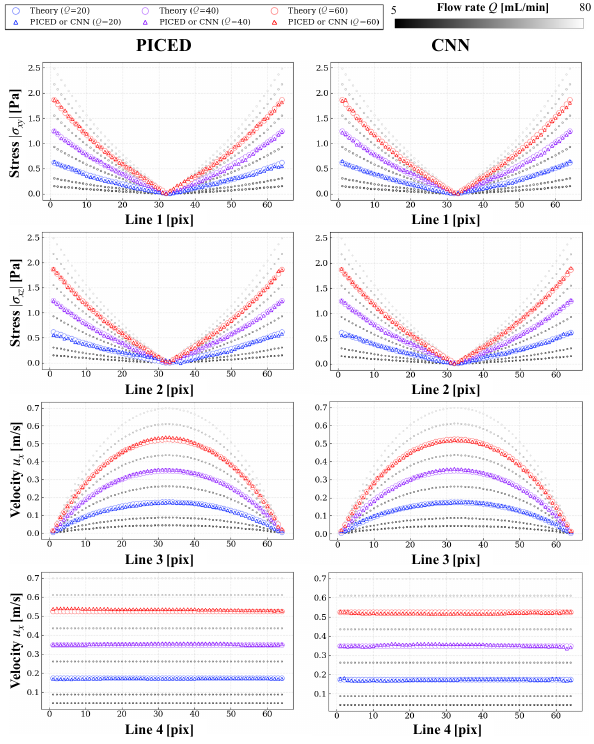}
\caption{Stress and velocity values for each line in Fig.~\ref{fig:MLresults40danmen} and for each flow rate. Gray markers represent the learning data, while colored markers indicate the testing data for $Q$ = 20, 40, and 60 mL/min. Circles denote theoretical values, and triangles represent the predictions obtained by PICED and CNN.}
\label{fig:MLresults_graph}
\end{figure*}

To deepen the discussion of the accuracy of the machine learning models, we present the line profiles of stress and velocity at the centerlines (Line 1:$x=z=32$ pix, Line 2:$x=y=32$ pix, Line 3:$x=z=32$ pix, Line 4:$y=z=32$ pix) for both PICED and CNN, as illustrated in Figure \ref{fig:MLresults_graph}.
The flow rates for the learning and testing data other than $Q=40$ mL/min are also included in the same figure.
Additionally, to enhance clarity, the absolute values of the stress are plotted in the graph.
It is evident from lines 1 to 4 in the figure that both PICED and CNN successfully achieve accurate predictions of the untrained interpolated data at almost all positions across all flow rates for the stress tensor components $\sigma_{xy}$,$\sigma_{xz}$, and the velocity $u_x$.

To further investigate the impact of incorporating physical equations in PICED, Table \ref{tab:PINN_CNN} presents the comparison of the error values represented by Equations (\ref{eq:loss}), (\ref{eq:LN}), and (\ref{eq:LC}) for the test results of PICED and CNN. 
Note that because the physical equations were not considered within one pixel from the wall in PICED, the errors were evaluated in regions excluding that pixel.
When comparing both models, there was no statistically significant difference between the predicted and true images with respect to the MSE $L_{\rm{data}}$.
On the other hand, the errors from the physical equations, $L_{\rm{N}}$ and $L_{\rm{C}}$, showed that PICED consistently had statistically significantly smaller errors than CNN across all flow rates.
Notably, for $L_{\rm{N}}$, as the flow rate increased, the error values for CNN increased, while those for PICED remained stable regardless of the flow rate.
This indicates that PICED successfully adhered more closely to the predictions based on physical equations.
Therefore, this study suggests that by utilizing machine learning, accurate predictions of the 3D stress distribution from integrated 2D images of phase difference and orientation can be achieved and that PICED is a more effective machine learning model for predictions that align with physical equations than CNN.

Additionally, to discuss the accuracy of the machine learning models, we calculated the relative squared error (RSE) for the stress, which is presented in Table \ref{tab:RSE}.
The RSE is defined by the following equation:

\begin{equation}
\label{eq:RSE}
\mathrm{RSE} = \frac{\sum\limits^N_{j=1} (A_j-P_j)^2}{\sum\limits^N_{j=1} (A_j-\bar{A})^2}.
\end{equation}

\noindent Table \ref{tab:RSE} demonstrates that the machine learning model developed in this study can predict stress with an accuracy of 2\% in terms of RSE.

\begin{table}[htbp]
\caption{Loss values of the data and the physical equations for CNN and PICED for each flow rate.}
\label{tab:PINN_CNN}
\begin{adjustwidth}{-10in}{-10in} \begin{center} \resizebox{0.5\textwidth}{!}{ 
\begin{tabular}{ccccccccc}
\hline\hline
& & $Q = 20$ mL/min & $Q = 40$ mL/min & $Q = 60$ mL/min\\
\hline
\multirow{3}{*}{CNN} & $L_{\rm{data}}$ & 1.05±0.30 [e-4] & 2.73±1.11 [e-4] & 1.40±0.45 [e-3]\\
& $L_{\rm{C}}$ & 1.78±0.68 [e-6] & 3.35±1.17 [e-6] & 4.94±2.67 [e-6]\\
& $L_{\rm{N}}$ & 6.06±2.03 [e-2] & 1.04±0.35 [e-1] & 2.01±1.28 [e-1]\\
\hline
\multirow{3}{*}{PICED} & $L_{\rm{data}}$ & 1.33±0.52 [e-4]	& 2.41±0.81 [e-4] & 1.08±0.26 [e-3]\\
& $L_{\rm{C}}$ & 8.10±2.08 [e-7] & 1.00±0.28 [e-6] & 1.43±0.53 [e-6]\\
& $L_{\rm{N}}$ & 1.19±0.35 [e-2] & 1.10±0.23 [e-2] & 1.03±0.07 [e-2]\\
\hline\hline
\end{tabular}
} \end{center} \end{adjustwidth} 
\end{table}

\begin{table}[htbp]
\caption{Relative squared errors (RSE) between the PICED predictions and the true stress values for each flow rate.}
\label{tab:RSE}
\begin{adjustwidth}{-10in}{-10in} \begin{center} \resizebox{0.5\textwidth}{!}{ 
\begin{tabular}{ccccccccc}
\hline\hline
& $Q = 20$ mL/min & $Q = 40$ mL/min & $Q = 60$ mL/min\\
\hline
$\sigma_{xy}$ (PICED) & 1.75±0.81 [e-2] & 4.56±1.23 [e-3] & 4.45±0.38 [e-3]\\
\hline
$\sigma_{xz}$ (PICED) & 1.33±0.11 [e-2] & 4.49±1.54 [e-3] & 3.99±0.83 [e-3]\\
\hline\hline
\end{tabular}
  } \end{center} \end{adjustwidth} 
\end{table}

\subsection{Component analysis}
This section presents an investigation of the effects of phase difference, orientation, and flow rate in the input images by examining the machine learning results when only the phase difference and when both the phase difference and orientation are used.
For simplicity of comparison, the model used was CNN, with all other conditions kept the same.
The MSE $L_{\rm{data}}$ for each test result is shown in Table \ref{tab:delta_phi}.
This table shows that there was no statistically significant difference between learning with only the phase difference and learning with both the phase difference and orientation.
Therefore, it was found that this model does not necessarily require orientation data.
Furthermore, when comparing the results of learning with both the phase difference and orientation as shown in Table \ref{tab:PINN_CNN} to those including the flow rate as presented in Table \ref{tab:delta_phi}, the $L_{\rm{data}}$ values for CNN were smaller when the flow rate was included.
Notably, for $Q=40$ mL/min, including the flow rate resulted in a statistically significant reduction in error, suggesting that including the flow rate in the input image enhances accuracy.
However, currently, the flow rate is simply input as an image that is the same size as the phase difference and orientation.
In the future, incorporating the flow rate values as boundary conditions could potentially lead to more robust predictions, particularly for values near the boundaries.

\begin{table}[t!]
\caption{Data loss of CNNs for cases in which both phase difference and orientation are learned, and in which only phase difference is learned in each flow rate.}
\label{tab:delta_phi}
\begin{adjustwidth}{-10in}{-10in} \begin{center} \resizebox{0.5\textwidth}{!}{ 
\begin{tabular}{ccccccccc}
\hline\hline
& & $Q = 20$ mL/min & $Q = 40$ mL/min & $Q = 60$ mL/min\\
\hline
$\Delta,\phi$ (CNN) & $L_{\rm{data}}$ & 2.02±0.70 [e-4] & 7.24±1.89 [e-4] & 1.66±0.64 [e-3]\\
\hline
$\Delta$ (CNN) & $L_{\rm{data}}$ & 2.35±0.62 [e-4] & 5.68±1.63 [e-4] & 1.05±0.48 [e-3]\\
\hline\hline
\end{tabular}
} \end{center} \end{adjustwidth} 
\end{table}

\subsection{Principal stress distribution}
This section describes the results of calculating the 3D maximum principal stress distribution acting within the fluid, using the 3D stress tensor distribution obtained earlier.
The data used for this calculation are the results shown in Figure \ref{fig:MLresults40}.
The maximum principal stress $\sigma_1$ and its direction vector $\boldsymbol{l}$ in the 3D rectangular channel flow are given by Equation (\ref{eq:sigma_3D}):

\begin{equation}
\label{eq:sigma_1}
\sigma_1 = \sqrt{\sigma_{xy}^2+\sigma_{xz}^2},
\end{equation}
\begin{equation}
\label{eq:psi}
\boldsymbol{l} =
\begin{bmatrix}
1\\
\frac{\sigma_{xy}}{\sqrt{\sigma_{xy}^2+\sigma_{xz}^2}}\\
\frac{\sigma_{xz}}{\sqrt{\sigma_{xy}^2+\sigma_{xz}^2}}
\end{bmatrix}.
\end{equation}

\noindent Here, atmospheric pressure $p$ was neglected.

\begin{figure}[htbp]
\centering
\includegraphics[width=1.0\columnwidth]{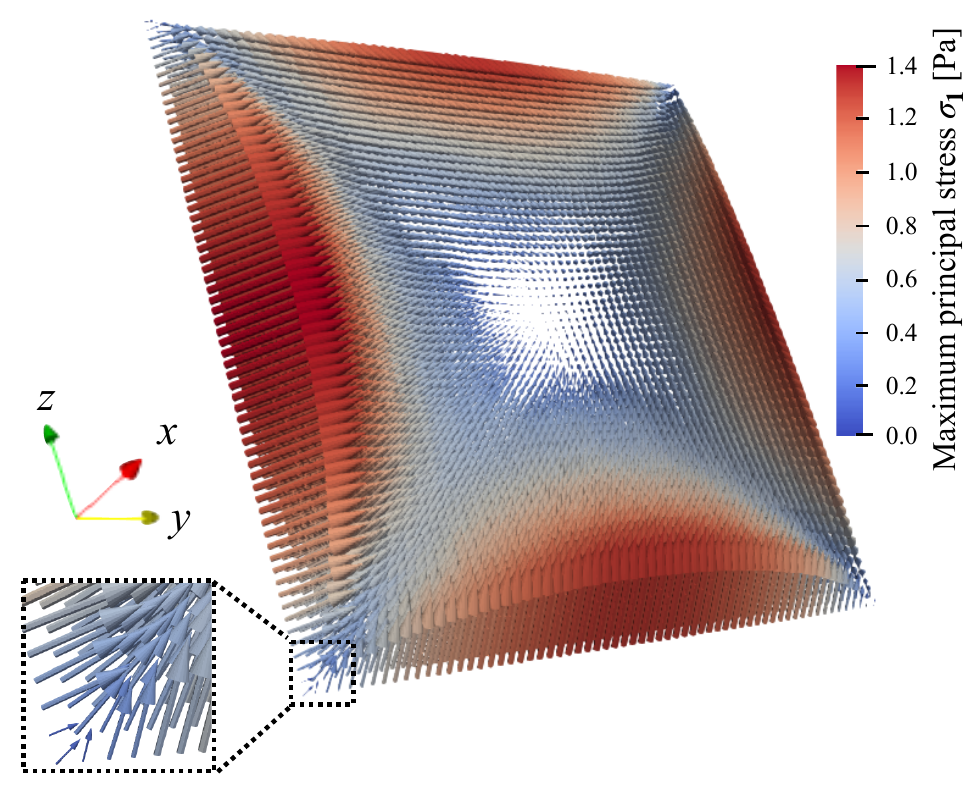}
\caption{3D map for the direction and the intensity of maximum principal stresses at $x$ = 32 pixels in $Q$ = 40 mL/min. The direction and length of the arrows represent the orientation and magnitude of the maximum principal stress, respectively.}
\label{fig:principal_single}
\end{figure}

The maximum principal stress distribution was calculated from the predicted stress tensor at cross-section ($x=32$ pix) shown in Figure \ref{fig:MLresults40}, and the results are depicted in Figure \ref{fig:principal_single}.
The length and color bars of the arrows represent the magnitude of the maximum principal stress, while their direction indicates its orientation.
The principal stress is observed to be large near the center of the four walls and decreases as it approaches the center of the flow path or the corners of the rectangular channel.
Furthermore, the direction of the principal stress points toward the flow path's center at all positions, indicating that shear stress is dominant.
These findings demonstrate that, for 3D flow in the rectangular channel, 3D stress tensor fields and principal stress distributions can be obtained by outputting interpolated data for multiple flow rates from the phase difference and orientation images acquired using photoelastic methods.

\section{Conclusion}
This study aimed to develop a machine learning model for obtaining a 3D stress field from 2D integrated images of phase difference and orientation captured by a polarizing camera.
This paper focused on 3D flow within a rectangular channel and conducted machine learning using a physics-informed convolutional encoder-decoder (PICED) that incorporates physical equations into the loss function to generate images from images.
We employed Cauchy’s equations of motion and the continuity equation, making the model applicable to fluid stress fields with unknown viscosity distributions.
As a result, PICED successfully predicted the 3D stress tensor distribution for multiple interpolation datasets with high accuracy, confirming that it has a smaller physical loss than CNN.
Additionally, by visualizing the distribution of maximum principal stress using the reconstructed stress tensor field, we achieved accurate predictions of both the location and direction of stress, demonstrating the model's potential as a fluid stress field analysis tool. 

Based on the results reported in this paper, we anticipate advancements in the use of images captured from multiple angles for 3D stress field reconstruction for complex fluids with unknown viscosity distributions.
Establishing these two methodologies could not only contribute directly to understanding the causes of aneurysms but also have a ripple effect across various fields that involve fluid dynamics and stress.

\section*{Acknowledgment}
This work was supported by Japan Society for the Promotion of Science (Grant No. 24H00289) and the Japan Science and Technology Agency (Grant No. PRESTO JPMJPR21O5). 
The authors would also like to thank Dr. Masaharu Kameda (Professor, Tokyo University of Agriculture and Technology) for helpful discussions and comments.

\bibliographystyle{elsarticle-num}
\footnotesize
\bibliography{Refs}

\end{document}